\documentclass{article}
\usepackage{spconf}

\usepackage{amsmath,amssymb,bbm}
\usepackage{color,graphicx}

\def\zZ{{\mathbb Z}}

\def\cost{{\mathcal C}}

\def\approx{\mbox{\footnotesize app}}
\def\opt{\mbox{\footnotesize opt}}
\def\appopt{\mbox{\footnotesize appopt}}
\def\subopt{\mbox{\footnotesize subopt}}
\def\worst{\mbox{\footnotesize wc}}

\title{Optimal Envelope Approximation in Fourier Basis with
Applications in TV White Space}
\name{Animesh Kumar}
\address{Department of Electrical Engineering\\ 
Indian Institute of Technology, Bombay\\ Mumbai, India -- 400076\\
animesh@ee.iitb.ac.in
}

\begin{document}

\maketitle

\begin{abstract}
Lowpass \textit{envelope approximation} of smooth continuous-variable signals are
introduced in this work. Envelope approximations are necessary when a given
signal has to be approximated always to a larger value (such as in TV white space
protection regions).  
In this work, a near-optimal approximate algorithm for finding a signal's
envelope, while minimizing a mean-squared cost function, is detailed. The sparse
(lowpass) signal approximation is obtained in the linear Fourier series basis.
This approximate algorithm works by discretizing the envelope property from an
infinite number of points to a large (but finite) number of points. It is shown
that this approximate algorithm is near-optimal and can be solved by using
efficient convex optimization programs available in the literature. Simulation
results are provided towards the end to gain more insights into the analytical
results presented.
% The convergence speed of the approximate algorithm towards the true optimum,
% in terms of the number of points in discretization of envelope property, is
% also established.
%
\end{abstract}

\section{Introduction}
\label{sec:intro}

\begin{figure}[!htb]
\centering
\includegraphics[scale=0.32]{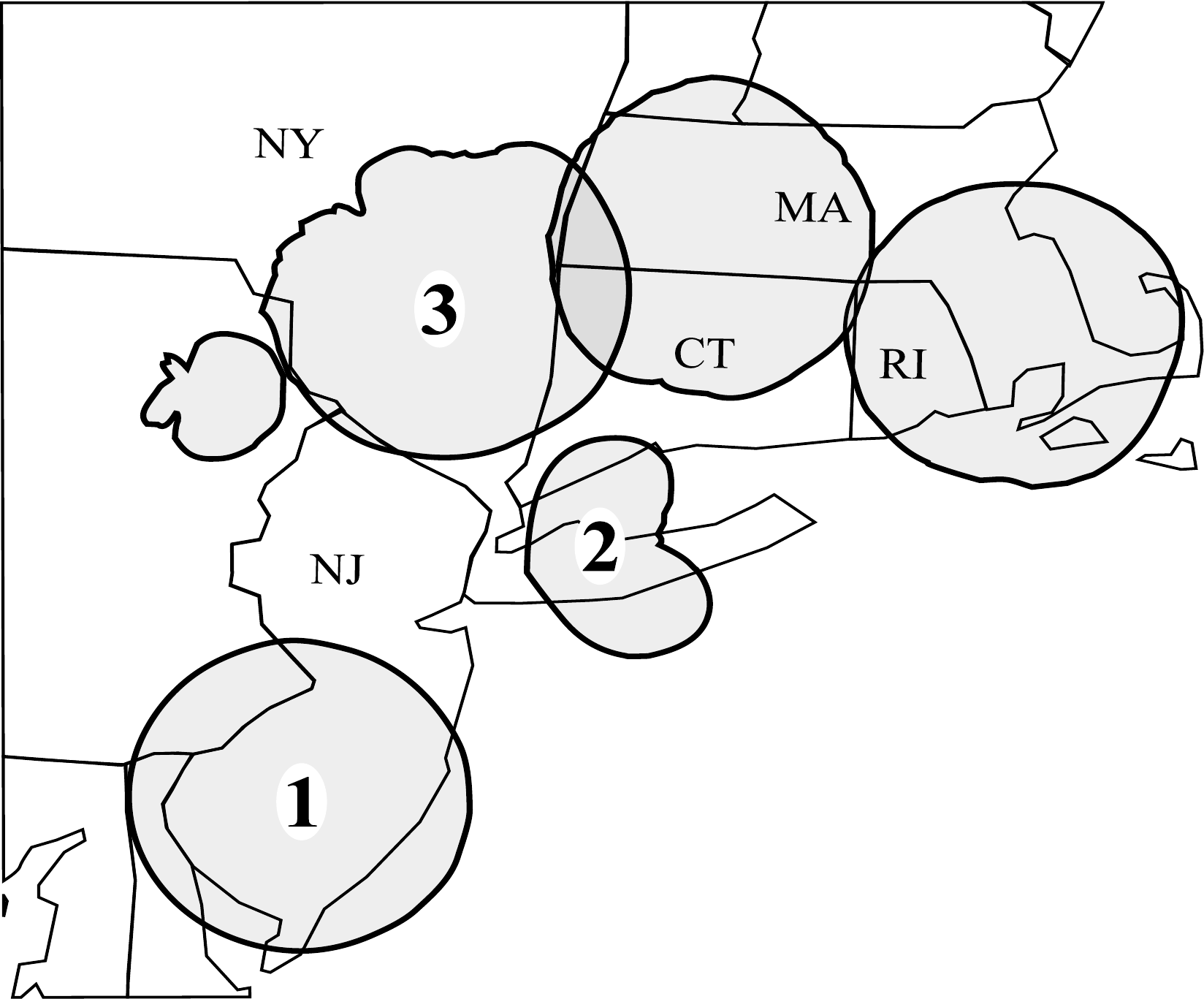}
\caption{\label{fig:coveragecontours} The TV protection regions for Channel~22
near New York, from the website of a TV white space service provider
\textit{iconectiv} in the United States~\cite{iconectivWEB}, are shown.
Protection regions labeled \textbf{2} and \textbf{3} are non-circular in shape.}
\end{figure}

This work introduces a \textit{fundamental topic} in some Electrical
Engineering applications---envelope approximations.  First, this problem is
motivated. TV white space devices are required to consult a TV white space
database~\cite{gurneyBEKGG2008}, which in turn computes the \textit{protection
region} of the TV transmitters. The TV white space database service providers
are licensed by a regulatory body such as the FCC in United States.  The
protection regions for the TV transmitters is smooth and can be non-circular in
shape; for example, see Fig.~\ref{fig:coveragecontours}, which illustrates
protection regions obtained from the iconectiv website~\cite{iconectivWEB} for
Channel~22 in the New York region. iconectiv is one of the database service
providers licensed by the FCC. Observe that protection regions such as
\textbf{2} and \textbf{3} are non-circular.  

The protection region signifies a closed region where only the licensed TV
transmitter can use the TV channel frequencies For example, in
Fig.~\ref{fig:coveragecontours}, in the region labeled \textbf{2} only licensed
user can operate in Channel~22 of the TV band.  If the TV white space database
wishes to communicate the protection region by using a lowpass (sparse or
rate-efficient) approximation, it needs to calculate a lowpass representation of
shapes such as \textbf{2} and \textbf{3}. While performing the approximation,
there are two possible errors: (i) a point in TV protection region is declared
as unprotected; and (ii) a point in unprotected region is declared as TV
protection region.  To protect the licensed operation of TV transmitters
type~(i) errors are \textit{not allowed} . So, any lowpass representation of TV
protection region must have an ``enveloping'' structure. To address such
problems, \textit{envelope approximations} are studied in this work.

\begin{figure}[!htb]
\centering
\includegraphics[scale=0.8]{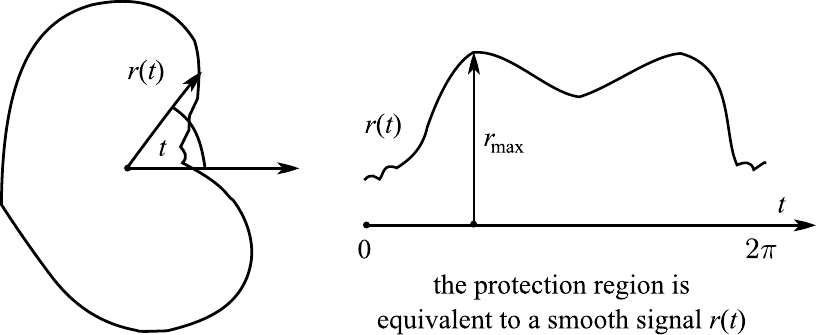}
\caption{\label{fig:better_approx} A protection region can be viewed as a
one-dimensional signal $r(t)$ with respect to the angle $t$ as shown.}
\end{figure}

Consider a smooth TV protection region as depicted in
Fig.~\ref{fig:better_approx}. Let its centroid be the origin.  Then, the
protection region can be \textit{parametrized} by a periodic signal $r(t)$ as
shown in Fig.~\ref{fig:better_approx}.  With the knowledge of origin (the
center), the signal $r(t)$ is equivalent to the protection region. This periodic
and smooth signal can be approximated by orthogonal basis in a linear space; for
example, Fourier series can be used~\cite{mallatSA2009}. In this work, $r(t)$
will be approximated to a bandlimited Fourier series $r_{\approx}(t)$,
where $r_{\approx}(t)$ has only $(L+1)$ harmonics in its Fourier series. The
envelope constraint requires $r_{\approx}(t) \geq r(t)$ for all $t \in [0,
2\pi]$, while minimizing a desirable \textit{cost function}. For TV protection
region approximation, the area enclosed by $r_{\approx}(t)$ should be minimized,
which means
% translates to the minimization of
%
\begin{align}
\mbox{minimize } & C(r_{\approx}, r) := \frac{1}{2} \int_0^{2\pi} r_{\approx}^2(t) \mbox{d} t
\nonumber \\
\mbox{ subject to } & r_{\approx}(t) \geq r(t). \label{eq:rcost}
\end{align}
This is the core problem addressed in this work.

\textit{Main result:} An \textit{approximate algorithm}, which banks upon convex
optimization program, is developed to address the optimization problem
in~\eqref{eq:rcost}. The approximate algorithm has two features: (i) it is
provably near-optimal to the best solution of optimization in~\eqref{eq:rcost};
and (ii) the nearness to optimality can be controlled by choosing the complexity
of solving the approximate algorithm.

\textit{Related work:} As far as we know, the topic of envelope approximation,
subject to a cost function, has not been addressed in the literature. This is a
fundamentally new topic. The topic of approximation or greedy approximation in
linear basis, on the other hand, is classically well
known~\cite{devoreLC1993,temlyakovG2011}.

\textit{Organization:} Section~\ref{sec:modeling} discusses the signal and
approximation model, and introduces the cost function.
Section~\ref{sec:iterative} presents the approximate algorithm for finding a
near-optimal envelope of a signal.  Section~\ref{sec:res} presents simulation
results while conclusions are in Section~\ref{sec:conclusions}.

\section{Modeling assumptions}
\label{sec:modeling}

A finite support real-valued field $f(t)$ will be considered, where $t \in
[0,1]$, without loss of generality.  It will be assumed that a periodic
repetition of $f(t)$, that is $\sum_{k \in \zZ} f(t -k)$, is differentiable in
$[0,1]$ so that Fourier basis is sparse for the
signal~\cite{mallatSA2009,devoreLC1993}. It will be assumed that
\begin{align}
|f'(t)| \leq c \label{eq:differentiable}
\end{align}
for some constant $c > 0$. With Fourier basis, the pointwise representation for
differentiable signals $f(t)$ is given by~\cite{mallatSA2009}
\begin{align}
f(t) = \sum_{ k \in \zZ} a[k] \exp{ (j 2\pi k t)} \label{eq:ffourier}
\end{align}
where the Fourier series coefficients are given by 
$
% \begin{align}
%
a[k] = \int_{0}^{1} f(t) \exp{(-j 2 \pi k t)} \mbox{d} t.
$
% \end{align}
%
% That is, % \begin{align} % f(t) = \sum_{ k = - \infty}^{\infty} a[k] \phi_k(t)
% \label{eq:fbasis} % \end{align} % where $\phi_k(t), k \in \zZ$ form the set of
% basis functions suitable for representing smooth signals with support in
% $[0,1]$. The set $\zZ$ denotes the set of integers.
Since $f(t)$ is real-valued,  conjugate symmetry implies $a[k] = \bar{a}[-k]$.
In general, to specify $f(t)$ completely, infinite number of coefficients
$\{a[0], a[1], a[2], \ldots \}$ have to be specified. 

In this work, an \textit{optimal envelope approximation} of $f(t)$ will be
designed. Let $f_{\approx}(t)$ be any $(L+1)$-complex coefficient based
envelope approximation.  Any $(L+1)$-coefficient envelope approximation will
have the following form:
\begin{align}
f_{\approx}(t) = \sum_{ k = - L}^{L} b[k] \exp{ (j 2\pi k t)},
\label{eq:envapprox}
\end{align}
where the envelope approximation will satisfy:
\begin{align}
f_{\approx}(t) \geq f(t) \quad \mbox{ for all } t \in [0,1].  \label{eq:envprop}
\end{align}
Since $f_{\approx}(t)$ is real-valued, the coefficients $b[k]$ and $b[-k]$ are
related by conjugate symmetry, that is $b[k] = \bar{b[-k]}$. The approximation
$f_{\approx}(t)$ is specified by $L+1$ coefficients $\{b[0], \ldots, b[L]\}$.
For compact notation let
\begin{align}
\vec{b} := \left( b[-L], b[-L+1], \ldots, b[L] \right)^T
\end{align}
where $\vec{b}$ is a column vector.

For the protection region approximation, a mean-squared cost will be minimized.
The cost is defined by
\begin{align}
\cost(f_{\approx}, f) & = \int_0^1 (f_{\approx}^2(t) - f^2(t))
\mbox{d}t
\end{align}
where $f_{\approx}(t)$ is the envelope approximation of $f(t)$. This cost 
represents the white space area lost as protection region.

\section{Optimal Envelope Approximation}
\label{sec:iterative}

This section presents a framework to obtain a near-optimal envelope
approximation for a smooth (differentiable) signal $f(t)$. The cost function is
assumed to be
\begin{align}
\cost(f_{\approx}, f) & = \int_0^1 (f_{\approx}^2(t) - f^2(t)) \mbox{d}t \\
& = \sum_{|k| \leq L} |b[k]|^2  - \sum_{i \in \zZ} |a[i]|^2 
\end{align}
where $f_{\approx}(t) \geq f(t)$.  Given a signal $f(t)$, its energy is fixed.
The envelope approximation problem is equivalent to finding
\begin{align}
\vec{b}_{\opt} & \in \arg\min_{\vec{b}}  \sum_{|k| \leq L} |b[k]|^2 \nonumber \\
\mbox{subject to } & f_{\approx}(t) \geq f(t) \quad \forall t \in [0,1].
\label{eq:optimization}
\end{align}
The signal $f(t)$ is fixed in the above optimization. For any fixed $t = t_0$
the constraint is linear since
\begin{align}
f_{\approx}(t_0) \geq f(t_0) \Leftrightarrow \vec{b}^T \vec{\phi}(t_0) \geq
f(t_0)
\end{align}
where $\vec{\phi}(t) := (\exp(- j 2 \pi L t), \exp(- j 2 \pi (L-1) t), \ldots,$
$\exp(j 2\pi Lt))^T$ is a vector of phasors. If the constraint
in~\eqref{eq:optimization}  was restricted to a finite number of points in
$[0,1]$, then the optimization in~\eqref{eq:optimization} can be solved as a
quadratic program with linear constraints.\footnote{The reader would notice that
$\vec{\phi}(t)$ is complex-valued, while quadratic program works with real
valued linear constraints. If $b[k] = b_R[k] + j b_I[k]$, then conjugate
symmetry implies that $b[-k] = b_R[k] - j b_I[k]$. These complex valued linear
constraints can be re-cast into real valued linear constraints in terms of
$b[0], b_R[1], b_I[1], \ldots, b_R[L], b_I[L]$. The details are omitted for
simplicity of the exposition and due to space constraints.} The quadratic
program is solvable using classical methods~\cite{boydVC2004}. So, the
\textit{difficulty} in solving the optimization in~\eqref{eq:optimization} is an
infinite number of constraints.

The signal $f(t)$ has been assumed to be differentiable. The
approximation $f_{\approx}(t)$ will be infinitely differentiable due to
bandlimitedness. This smoothness of $f_{\approx}(t) - f(t)$ suggests  that if
$f_{\approx}(t_0) - f(t_0) \geq 0$, it will be positive or near-zero in a small
interval around $t_0$. This intuition motivates an $n$-point approximation to
the constraint of optimization problem in~\eqref{eq:optimization}.
% This is discussed next.
Consider the following optimization, which is an $n$-point
approximation to the optimization in~\eqref{eq:optimization}
\begin{align}
\vec{b}_{\appopt,n} & = \arg\min_{\vec{b}} \sum_{|k| \leq L} |b[k]|^2
\nonumber \\
\mbox{subject to } f_{\approx}(t) & \geq f(t) \quad \forall t \in
\left\{0, \frac{1}{n}, \ldots,  \frac{n-1}{n}\right\}.
\label{eq:approxoptimization}
\end{align}
The above optimization program has a quadratic cost with $n$ linear
constraints and it is solvable by a convex program
solver~\cite{boydVC2004}.  Let $\vec{b}_{\opt}$ and
$\vec{b}_{\appopt}$ be the unique arguments for
which~\eqref{eq:optimization} and~\eqref{eq:approxoptimization} are
minimized. Then $\vec{b}_{\appopt,n}$ can be solved with a convex
program. It is expected, though unproved so far, that $\vec{b}_{\opt}$
and $\vec{b}_{\appopt,n}$ will be ``close'' as $n$ becomes large. Their
closeness is established next.

First note that there are more constraints in~\eqref{eq:optimization}
than in~\eqref{eq:approxoptimization}. Therefore,
\begin{align}
\sum_{|k| \leq L} |b_{\opt}[k]|^2 \geq \sum_{|k| \leq L}
|b_{\appopt,n}[k]|^2. \label{eq:increasing}
\end{align}
A sub-optimal approximation $f_{\subopt,n}(t)$ with Fourier
series $\vec{b}_{\subopt,n}$ will be constructed using
$\vec{b}_{\appopt,n}$ such that $f_{\subopt,n}(t) \geq f(t)$ for all
$t \in [0,1]$. 
Assume that $|f_{\appopt,n}'(t)| \leq c'$, where $c'$ is a finite
constant, which is proved later in this section. Since
$f_{\appopt,n}(t)$ is obtained by solving optimization
in~\eqref{eq:approxoptimization}, therefore
\begin{align}
f_{\appopt,n}\left( \frac{i}{n} \right) \geq f\left( \frac{i}{n}
\right), \forall i \in \{0, 1, \ldots, n-1\} \label{eq:fappibyn}
\end{align}
because of constraint equation. For any point $t \in [\frac{i}{n},
\frac{i+1}{n}]$
\begin{align}
& f_{\appopt,n}(t) - f(t) \nonumber \\ 
& \stackrel{(a)}{\geq} f_{\appopt,n}\left( \frac{i}{n} \right) - c'
\left( t - \frac{i}{n} \right) - \left[ f\left( \frac{i}{n} \right) +
c \left( t - \frac{i}{n} \right) \right] \nonumber \\
& \stackrel{(b)}{\geq} f_{\appopt,n}\left( \frac{i}{n} \right) -
f\left( \frac{i}{n} \right)  - \frac{c + c'}{n} \\
& \stackrel{(c)}{\geq} - \frac{c + c'}{n} 
\end{align}
where $(a)$ follows by $f_{\appopt,n}(t) \geq f_{\appopt,n}(t_0) -
c'(t - t_0)$ and $f(t) \leq f(t_0) + c (t - t_0)$ for $t \geq t_0$,
$(b)$ follows by $t - i/n \leq 1/n$ for $t \in [\frac{i}{n},
\frac{i+1}{n}]$, and $(c)$ follows by~\eqref{eq:fappibyn}. The above
inequality holds for every $i$ (uniformly), so 
\begin{align}
f_{\appopt,n}(t) - f(t) \geq - \frac{c + c'}{n} \quad \forall t \in
[0,1]. \label{eq:appoptbound}
\end{align}
Define
\begin{align}
f_{\subopt,n}(t) := f_{\appopt,n}(t) + \frac{c + c'}{n}
\label{eq:suboptdefn}
\end{align}
which means 
\begin{align}
b_{\subopt,n}[k] & = b_{\appopt,n}[k] \quad \mbox{for } k \neq 0
\nonumber \\
& = b_{\appopt,n}[0] + \frac{c +c'}{n} \quad \mbox{for } k = 0
\label{eq:bsubopt}
\end{align}
From~\eqref{eq:appoptbound} and \eqref{eq:suboptdefn}, it follows that
\begin{align}
f_{\subopt,n}(t) \geq f(t)
\end{align}
or $f_{\subopt}(t)$ satisfies the constraint
in~\eqref{eq:optimization}, which means
\begin{align}
\sum_{|k| \leq L} |b_{\opt}[k]|^2 \leq \sum_{|k| \leq L}
|b_{\subopt,n}[k]|^2. \label{eq:decreasing}
\end{align}
From~\eqref{eq:increasing} and \eqref{eq:decreasing}, we get the
following key inequality
\begin{align}
\sum_{|k| \leq L} |b_{\appopt,n}[k]|^2 \leq \sum_{|k| \leq L}
|b_{\opt}[k]|^2  \leq \sum_{|k| \leq L} |b_{\subopt,n}[k]|^2.
\nonumber 
\end{align}
By this inequality, 
\begin{align}
\sum_{|k| \leq L} & |b_{\opt}[k]|^2 - |b_{\appopt,n}[k]|^2 \\
& \leq \sum_{|k| \leq L} |b_{\subopt,n}[k]|^2 - |b_{\appopt,n}[k]|^2
\\
& = \left| b_{\appopt,n}[0] + \frac{c + c'}{n} \right|^2 -
|b_{\appopt,n}[0]|^2  \\
& = 2 b_{\appopt,n}[0]  \frac{c + c'}{n} +  \left(\frac{c +
c'}{n}\right)^2 \\
& = 2  b_{\appopt,n}[0]  \frac{c + c'}{n} + o(1/n).
\label{eq:appoptaccuracy}
\end{align}
Similarly,
\begin{align}
\sum_{|k| \leq L} & |b_{\subopt}[k]|^2 - |b_{\opt}[k]|^2 \\
& \leq \sum_{|k| \leq L} |b_{\subopt,n}[k]|^2 - |b_{\appopt,n}[k]|^2
\\
& = 2  b_{\appopt,n}[0]  \frac{c + c'}{n} + o(1/n).
\label{eq:suboptaccuracy}
\end{align}
The above results guarantee that the cost obtained by
$f_{\subopt,n}(t)$, a suboptimal solution to the envelope
approximation problem obtained through $f_{\appopt,n}(t)$, is
\textit{at-most} $O(1/n)$ away from the true optimum. If $n$ is
large-enough, the above discussion guarantees that \textit{a
near-optimal solution to the envelope approximation problem can be
obtained} in an efficient way.

It remains to show that $|f_{\appopt,n}'(t)| \leq c'$. First note that
\begin{align}
|f_{\appopt,n}'(t)| & = \left| \sum_{|k|\leq L} j 2 \pi k
b_{\appopt,n}[k] \exp(j 2 \pi kt) \right| \\
& \leq   \sum_{|k|\leq L} 2 \pi |k b_{\appopt,n}[k]| \\
& \leq \sum_{|k| \leq L} 2 \pi L |b_{\appopt,n}[k]| \\
& \leq 2 \pi L \left[ \sum_{|k| \leq L} |b_{\appopt,n}[k]|^2
\right]^{1/2} \label{eq:intermslope}
\end{align}
Next, note that $f_{\worst}(t) \equiv \|f\|_\infty$
is always a part of the constraint set in~\eqref{eq:optimization} and
\eqref{eq:approxoptimization}. Therefore,
\begin{align}
\sum_{|k| \leq L} |b_{\appopt,n}[k]|^2 & \leq \int_{0}^1
|f_{\worst}(t)|^2 \mbox{d} t
= \|f\|_\infty^2. \label{eq:costbound}
\end{align}
Substitution of~\eqref{eq:costbound} in~\eqref{eq:intermslope} results
in
\begin{align}
c':= | f_{\appopt, n}'(t) | \leq & 2 \pi L  \|f\|_\infty
\label{eq:slope}
\end{align}
This concludes the proof of $c' < \infty$. Simulations are presented next.

\section{Simulations on TV protection region}
\label{sec:res}

\begin{figure}[!htb]
\centering
\includegraphics[scale=0.40]{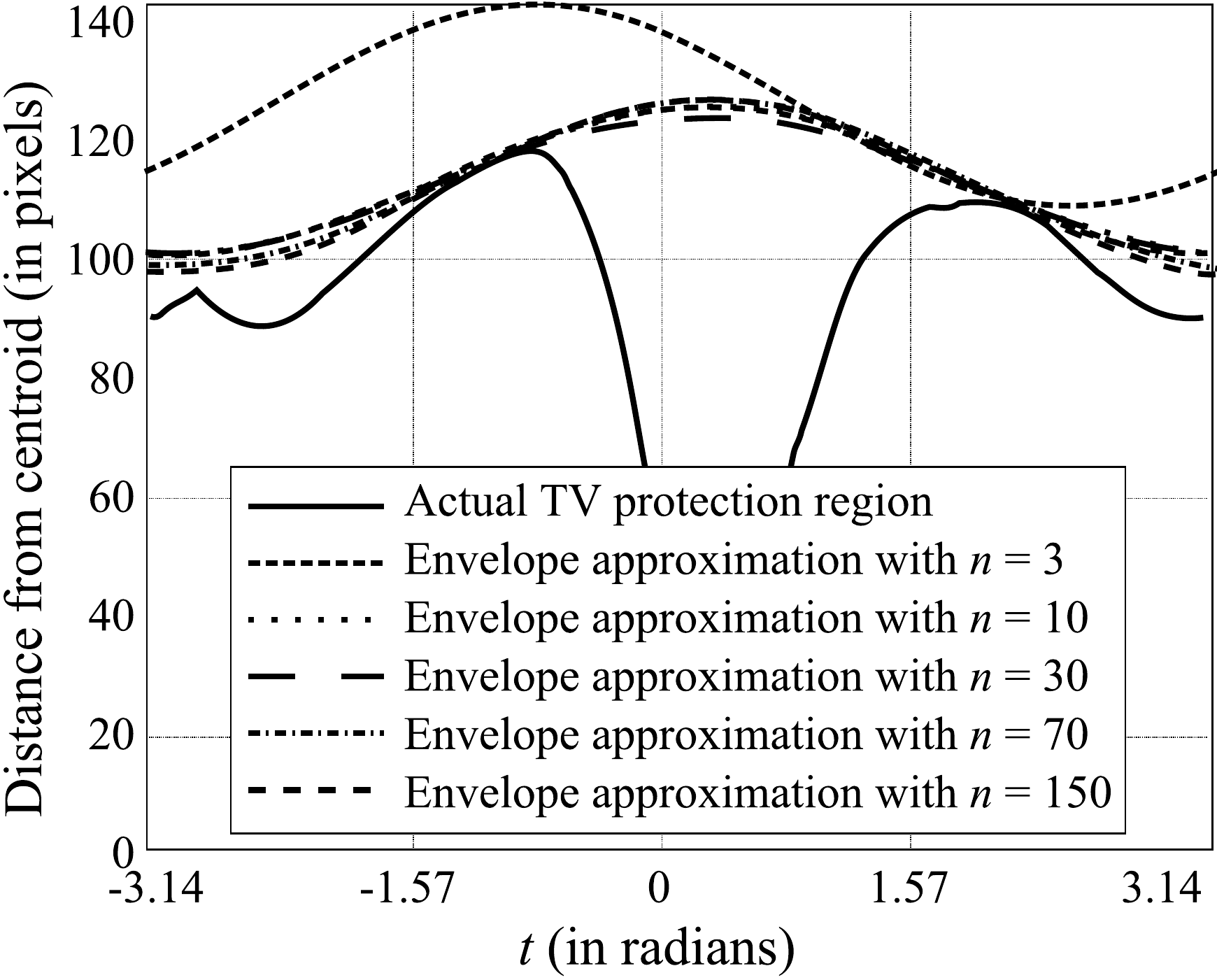}
\caption{\label{fig:changingN} The sub-optimal envelope approximation
$r_{\subopt,n}(t)$ is obtained as a function of $n$. Here $L = 1$. It is
observed that $r_{\subopt,n}(t)$ is proximal to optimal for $n \geq 10$.
By design, the approximations are larger than $r(t)$ for each value of $t$.}
\end{figure}

\begin{figure}[!htb]
\centering
\includegraphics[scale=0.40]{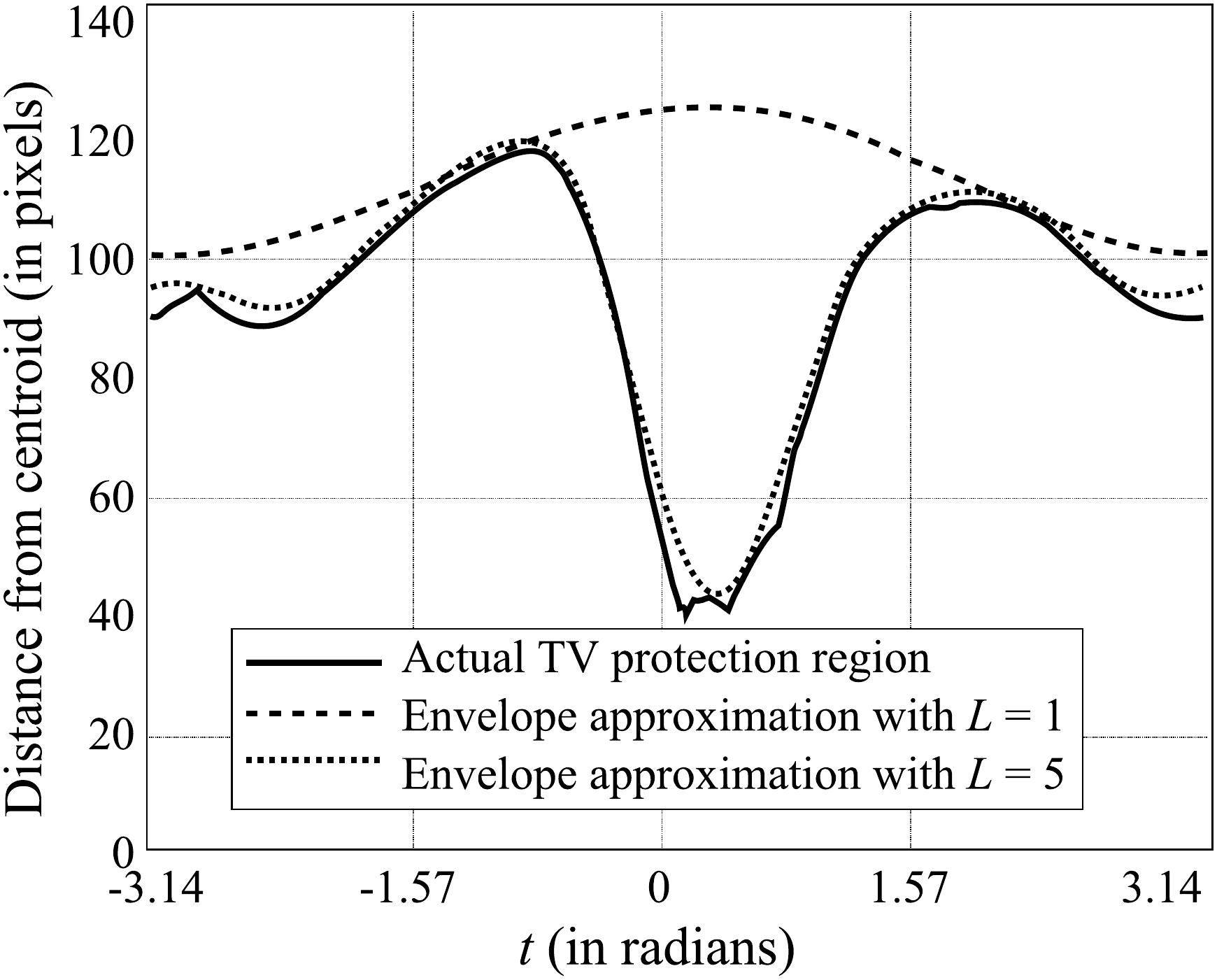}
\caption{\label{fig:changingL} The optimal envelope approximation, a solution to
\eqref{eq:optimization}, is illustrated for $L = 1$ and $L = 5$.  By design, the
approximation is larger than $r(t), t \in [-\pi, \pi]$.  This property also
ensures that the approximate protection region is a superset of the actual
protection region.}
\end{figure}

To test the optimal envelope approximation method of the previous section, TV
protection regions  in Channel~$2$ of United States were examined using the TV
white spaces US Interactive Map of Spectrum Bridge.  Across United States, there
are $57$ protected service contours. One of these contours was hand-picked and
its protection region was segmented (using image processing techniques) to
obtain $r(t), t \in [-\pi, \pi]$. This protection region was picked since it has
points with sudden change in derivative, and would be difficult to approximate.
Then, the Fourier basis based envelope approximation technique was applied
(see~\eqref{eq:bsubopt}).  The results are shown in Fig.~\ref{fig:changingN} and
Fig.~\ref{fig:changingL}. In Fig.~\ref{fig:changingN}, $L = 1$ and $n$ is
increased from $3$ onwards to obtain $r_{\subopt,n}(t)$. It is observed that
$r_{\subopt,n}(t)$ nearly converges for $n \geq 10$.  In
Fig.~\ref{fig:changingL}, $L = 1$ is increased to $L = 5$. It is observed that
with $11$ Fourier series coefficients, the envelope is proximal to the original
signal $r(t)$, except near derivative discontinuity.

\section{Conclusions}
\label{sec:conclusions}

An approximate algorithm for finding a signal's envelope, while minimizing a
mean-squared cost function, was detailed. A near-optimal envelope approximation
was found in Fourier basis using linear space properties and efficient
solvability of quadratic optimization subject to linear
constraints. The approximate algorithm when subjected to $n$-constraints 
resulted in a near-optimal envelope signal with a gap of $O(1/n)$ in
the cost function from the optimum. The results were verified with simulations
on TV white space protection region.

% \section{Acknowledgments}

\bibliographystyle{IEEEbib}

\end{document}